\documentclass{aa}  
\usepackage{graphicx}
\usepackage{txfonts}
\begin{document}
   \title{The $R_h=ct$ Universe Without Inflation}

   \author{F. Melia
%          \inst{0}
\fnmsep\thanks{John Woodruff Simpson Fellow}
          }

   \offprints{F. Melia}

   \institute{Department of Physics, The Applied Math Program, and Department of Astronomy, 
The University of Arizona, Tucson, Arizona 85721, USA \\
\email{fmelia@email.arizona.edu}
          }

   \date{Received September 26, 2012}

% \abstract{}{}{}{}{} 
% 5 {} token are mandatory
 
  \abstract
  % context heading (optional)
  % {} leave it empty if necessary  
   {The horizon problem in the standard model of cosmology ($\Lambda$DCM)
arises from the observed uniformity of the cosmic microwave background radiation,
which has the same temperature everywhere (except for tiny, stochastic fluctuations),
even in regions on opposite sides of the sky, which appear to lie outside of each other's
causal horizon. Since no physical process propagating at or below lightspeed could have
brought them into thermal equilibrium, it appears that the universe in its infancy required
highly improbable initial conditions. } 
  % aims heading (mandatory)
   { In this paper, we demonstrate
that the horizon problem only emerges for a subset of FRW cosmologies, such as
$\Lambda$CDM, that include an early phase of rapid deceleration.}
  % methods heading (mandatory)
   { The origin of the problem is examined by
considering photon propagation through a Friedmann-Robertson-Walker (FRW)
spacetime at a more fundamental level than has been attempted before.}
  % results heading (mandatory)
   { We show that
the horizon problem is nonexistent for the recently introduced $R_{\rm h}=ct$ universe,
obviating the principal motivation for the inclusion of inflation. We
demonstrate through direct calculation that, in this cosmology,
even opposite sides of the cosmos have remained causally connected to us---and
to each other---from the very first moments in the universe's expansion. Therefore,
within the context of the $R_{\rm h}=ct$ universe, the hypothesized inflationary
epoch from $t=10^{-35}$ seconds to $10^{-32}$ seconds was not needed to fix
this particular ``problem,'' though it may still provide benefits to cosmology for
other reasons.}
  % conclusions heading (optional), leave it empty if necessary 
   {}


   \maketitle
%
%________________________________________________________________

\section{Introduction}
The horizon problem arises out of the standard model of cosmology,\footnote{Today,
the standard model of cosmology ($\Lambda$CDM) is taken to be a particular solution to Einstein's
equations of general relativity, in which the overall energy density is comprised of
three principal components: (luminous and dark) matter, radiation, and an as yet unidentified
dark energy, which is usually assumed to be a cosmological constant $\Lambda$.}
which posits that the universe began with a hot big bang, followed by an expansion
calculable from the solution to two rather simple differential
equations. In this picture, the cosmos we see today began its existence
as a hot, dense plasma of thermalized matter and radiation, subjugated
dynamically by its own self-gravity and the possible effects of an
unspecified dark energy, whose principal role is to bring about
the apparent acceleration we see today (Riess \cite{Riess98}; 
Perlmutter \cite{Perlmutter99}). The initial hot state was followed 
by a gradual cooling and a drop in density and energy.

The observational evidence for a ``beginning'' appears to be overwhelming.
Distant galaxies are receding from us at a speed proportional to their
distance. A pervasive, relic microwave background radiation has been
discovered with a Planck spectrum corresponding to a (highly
diminished) temperature of $2.7^\circ$ K. And the measured cosmic
abundances are in agreement with the ratios predicted by big-bang
nucleosynthesis, which produced elements with a composition of
roughly $75\%$ hydrogen by mass, and about $25\%$ helium.

Couched in the language of general relativity, these observations
confirm certain assumed symmetries implied by the Cosmological Principle,
which holds that the universe is homogeneous and isotropic on large scales,
certainly larger than 100 Mpc. The differential equations describing the
universal expansion emerge directly from Einstein's theory when these
symmetries are adopted. They allow several different kinds of behavior
as the constituents vary, and account very well for the global
characteristics seen from the moment of the big bang all the way to the
present time $t_0$, some 13.7 billion years later.

The horizon problem is viewed as a major shortcoming of the standard model---not
as a result of an observational or internal inconsistency but, rather, from the
fact that the universe seems to have required special initial conditions that
are highly improbable according to our current understanding of physics. In its
most direct manifestation, the horizon problem arises from the observed
uniformity of the microwave background radiation, which has the same
temperature everywhere, save for fluctuations at the level of one part in
100,000 seen in WMAP's measured relic signal (Spergel \cite{Spergel03}). Regions
on opposite sides of the sky, the argument goes, lie beyond each other's
horizon (loosely defined as the distance  light could have traveled during 
a time $t_0$), yet their present temperature is identical, even though they could
not possibly have ever been in thermal equilibrium, since no physical process
propagating at or below the speed of light could have causally connected them.
The horizon problem therefore emerges as the inability of the standard model
to account for this homogeneity on scales greater than the distance light
could have traveled since the big bang, requiring some highly tuned spatial
distribution of temperature that has no evident natural cause.

So serious has this shortcoming become that the inflationary model of
cosmology (Guth \cite{Guth81,Linde82}) was invented in part to resolve this possible
discrepancy. In this picture,
an inflationary spurt occurred from approximately $10^{-35}$ seconds to
$10^{-32}$ seconds following the big bang, forcing the universe to expand
much more rapidly than would otherwise have been feasible solely under
the influence of matter, radiation, and dark energy, carrying causally
connected regions beyond the horizon each would have had in the
absence of this temporary acceleration.

In this paper, we will analyze the horizon problem more carefully than has
been attempted before, demonstrating that the conflict between
the observed properties of the cosmic microwave background (CMB) and
$\Lambda$CDM is not generic to all FRW cosmologies. In particular, we
will show that in the $R_{\rm h}=ct$ universe, even opposite sides of the
cosmos have remained causally connected to us---and to each other---from
the very first moments of the expansion. In this cosmology, therefore,
the inflationary model may be attractive for other reasons, but there is
actually no real horizon problem for it to resolve. We will argue that
the inconsistency between the observed properties of the CMB
and $\Lambda$CDM could be an indication that the standard model
requires a new ingredient---inflation---to fix it, but could also constitute
evidence in support of the $R_{\rm h}=ct$ universe.

\section{The $R_{\rm h}=ct$ Universe}

It will be helpful in what follows for us to briefly review what we know about 
the $R_{\rm h}=ct$ universe up to this point. From a theoretical standpoint, 
the $R_{\rm h}=ct$ condition is required when one adopts both the Cosmological 
principle and Weyl's postulate together (Melia \& Shevchuk \cite{MeliaShevchuk12}). 
If it turns out that the universe is indeed not correctly described by the 
$R_{\rm h}=ct$ cosmology, then either the Cosmological principle or (more 
likely) Weyl's postulate would be called into question. That too would be quite 
interesting and profoundly important, but beyond the scope of our present 
discussion. (A more pedagogical description of the $R_{\rm h}=ct$ universe
is given in Melia 2012a.)

Over the past several decades, $\Lambda$CDM has developed into a
comprehensive description of nature, in which a set of primary constituents
(radiation, matter, and an unspecified dark energy) are assumed with a partitioning 
determined from fits to the available data. $\Lambda$CDM must therefore 
deal with the challenges of accounting for varied observations covering disparate
properties of the universe, some early in its history (as seen in the relic CMB)
and others more recently (such as the formation of large-scale structure).

It is therefore desirable to keep testing the basic premisses of the theory,
particularly those that have the most profound consequence, such as the
possible role that inflation may have played in the early universe.
In recent papers, we have demonstrated that the $R_{\rm h}=ct$ universe 
may be able to provide a physical basis for the features
now emerging from the observations. For example, whereas one must infer
empirically the overall equation-of-state of the constituents in $\Lambda$CDM, 
the total pressure $p$ in $R_{\rm h}=ct$ is given by the simple expression 
$p=-\rho/3$, in terms of the total energy density $\rho$. This leads
to a great simplification of the observable quantities, such as the luminosity 
distance and the redshift-dependent Hubble constant, both of which
appear to be consistent with the data (see, e.g., Melia \& Maier 2013;
Wei et al. 2013). A more comprehensive description of the $R_{\rm h}=ct$
universe, and tests of its predictions against the observations, may
be found elsewhere (see, e.g., Melia 2012b, 2013; Melia \& Maier 2013;
Wei et al. 2013). In the context of this paper, the $R_{\rm h}=ct$
cosmology provides a good counterpoint to $\Lambda$CDM because,
unlike the latter, the former does not require a period of inflation.

\section{A Simple Analogy In Flat Spacetime}

Let us now begin our discussion of the horizon problem by considering
a much simpler analogy in flat spacetime,
in which two frames of reference, A and C, coincident with ours (frame B)
at time $t=0$ (with $t$ measured in our frame), are receding from us at speed
$v$  (see Figure~1). If $v$ is close to $c$, inevitably we reach the point where
$2vt>ct$, and we would naively conclude that A and C are beyond
each other's ``horizon," here defined simply as $ct$. But there
are actually several reasons why this statement is incorrect.

First, we cannot use the coordinates in our frame B to determine who
is causally connected to A or C by simply
adding velocities. To be specific, let's assume that $v=0.9c$. Then,
A's speed relative to C is not $2v=1.8c$ but, by the relativistic
addition of velocities, A is moving at speed $0.994475c$ relative to C,
and is therefore a distance $0.994475ct'$ ($< ct'$) away from C
as measured after a time $t'$ in C's frame. In other words, there
is no violation of causality when A's distance is measured by the
observer in C using his coordinates, rather than those of another
observer (in this case, ours) in frame B.

   \begin{figure}[hp]
   \centering
      \includegraphics[angle=0,width=8.3cm]{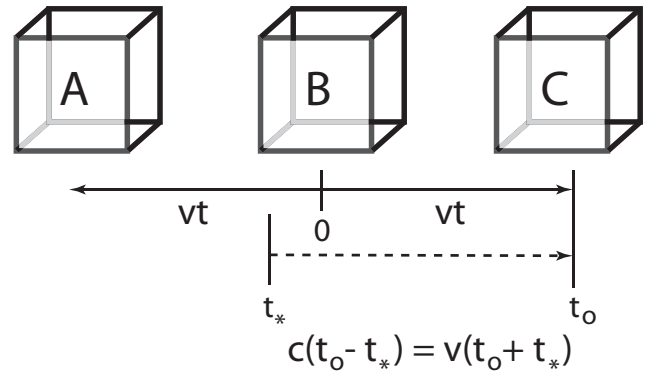}
      \caption{Simple analogy in flat spacetime, in which 2 frames of
reference (A and C) are moving away from us (situated in frame B), each
at speed $v$, though in opposite directions. If $v$ is close to $c$,
after a time $t$ in our frame, A and C are apparently beyond each
other's horizon (here defined as $ct$), but A and C could still have
communicated after $t=0$ (and before $t_0$) if light were emitted by
A at time $t_*$, such that $c(t_0-t_*)=v(t_0+t_*)$.}
\end{figure}

The second reason is that even though eventually
$2vt>ct$, we would still conclude that A and C had communicated
with each other after $t=0$, as long as they initiated the process
very soon after separation had begun. Suppose we see A sending a light
signal to C at time $t_*$. We would conclude that the light signal had
reached C by the time $t_0$, if
\begin{equation}
c(t_0-t_*)=v(t_0+t_*)
\end{equation}
or, in other words, we would see that A and C had communicated
with each other if their signals had been sent prior to the time
\begin{equation}
t_*\approx {1\over 2\gamma^2}t_0\;,
\end{equation}
where $\gamma\equiv (1-[v/c])^{-1/2}$  is the Lorentz factor.
The emission time $t_*$ is small, but not zero. So regardless of whether
we argue from our perspective in B, or from that of the two other
frames, all observers would agree that A and C had been in causal
contact after $t=t'=0$, as long as all three frames were coincident at the
beginning. This is true in spite of the fact that $2vt_0>ct_0$ in our frame.
As we shall see shortly, this example is quite instructive because an effect
similar to this emerges when we talk about proper distances between
us and the various patches of the CMB on opposite sides of the sky (see Figure~2 below).

\section{Null Geodesics in the FRW Spacetime}

The Friedmann-Robertson-Walker (FRW) metric for a spatially homogeneous
and isotropic three-dimensional space may be written in terms of the cosmic
(or proper) time $t$ measured by a co-moving observer, and the corresponding
radial ($r$) and angular ($\theta$ and $\phi$) coordinates in the co-moving frame,
for which  the interval $ds$ is
\begin{equation}
ds^2 = c^2 dt^2 - a^2(t)\left[{dr^2\over (1 - kr^2)} + r^2(d\theta^2 + 
\sin^2\theta d\phi^2)\right]\;.
\end{equation}
The constant $k$ is $+1$ for a closed
universe, $0$ for a flat, open universe, or $-1$ for an open universe. The spatial coordinates
in this frame remain ``fixed" for all particles in the cosmos, while their physical separation
varies by the scale factor $a(t)$. And to provide a framework for the analysis that
follows, let us point out that, for radial motion,
$a(t)dr/(1-kr^2)$ is the proper {\it incremental} distance realized when
$dt=0$, but only for each specific observer at his/her location. Nowadays,
it is common for people to refer to the global quantity
\begin{equation}
R(t)\equiv a(t)\int {dr\over\sqrt{1-kr^2}}
\end{equation}
as the proper distance, but one must remember that $R(t)$ is calculated assuming
the same common time $t$ everywhere, which is {\it not} the physical time for that
observer at locations other than at $r=0$ (more on this below). Understanding this
distinction will be critical for a full appreciation of why these co-moving coordinates
often lead to confusion regarding what is---and what is not---causally connected.

Utilizing the FRW metric in Einstein's field equations of general relativity, one obtains
the corresponding FRW differential equations of motion alluded to in the introduction.
These are the Friedmann equation,
\begin{equation}
H^2\equiv\left({\dot a\over a}\right)^2={8\pi G\over 3c^2}\rho-{kc^2\over a^2}\;,
\end{equation}
and the ``acceleration" equation,
\begin{equation}
{\ddot a\over a}=-{4\pi G\over 3c^2}(\rho+3p)\;.
\end{equation}
An overdot denotes a derivative with respect to cosmic time $t$, and $\rho$ and $p$
represent the total energy density and total pressure, respectively. A further application
of the FRW metric to the energy conservation equation in general relativity yields the final
equation,
\begin{equation}
\dot\rho=-3H(\rho+p)
\end{equation}
which, however, is not independent of Equations~(5) and (6).

Using the definition of proper radius, we can now derive the differential equation for
photon trajectories in a cosmology consistent with the FRW metric (Equation~3). We have
\begin{equation}
\dot{R}=\dot{a}r+a\dot{r}\;,
\end{equation}
and the null condition applied to Equation~(3) yields
\begin{equation}
c\,dt=-a(t){dr\over\sqrt{1-kr^2}}\;,
\end{equation}
where we have assumed propagation of the photon along a radius
towards the origin. The best indications we have today are
that the universe is flat so, for simplicity, we will assume
$k=0$ in all the calculations described below. Therefore,
$\dot{r}=-c/a$ for a photon approaching the origin, and
Equation~(8) becomes
\begin{equation}
\dot{R}_\gamma=c\left({R_\gamma\over R_{\rm h}}-1\right)\;,
\end{equation}
where we have added a subscript $\gamma$ to emphasize the
fact that this represents the proper radius of a photon
propagating towards the observer. This expression makes use of
the gravitational (or Hubble) radius (Melia \cite{Melia07}; Melia \& Shevchuk \cite{MeliaShevchuk12})
\begin{equation}
R_{\rm h}= {c\over H(t)}\;,
\end {equation}
and we note that
both $R_\gamma$ and $R_{\rm h}$ are functions of cosmic time $t$.
When the equation of state is written in the form $p=w\rho$ (for the
total pressure $p$ in terms of the total energy density $\rho$),
it is not difficult to show from the Friedmann (5) and acceleration (6)
equations that the gravitational radius satisfies the dynamical
equation (Melia \& Abdelqader \cite{MeliaAbdelqader09})
\begin{equation}
\dot{R}_{\rm h}={3\over 2}(1+w)c\;.
\end{equation}
Solving Equations~(10) and (12)
simultaneously yields the null geodesic
$R_\gamma(t)$ linking a source of photons at proper
distance $R_{\rm src}(t_e)=R_\gamma(t_e)$ and cosmic time $t_e$,
with the observer who receives them at time $t_0$, when
$R_\gamma(t_0)=0$.

   \begin{figure}[hp]
   \centering
      \includegraphics[angle=0,width=8.3cm]{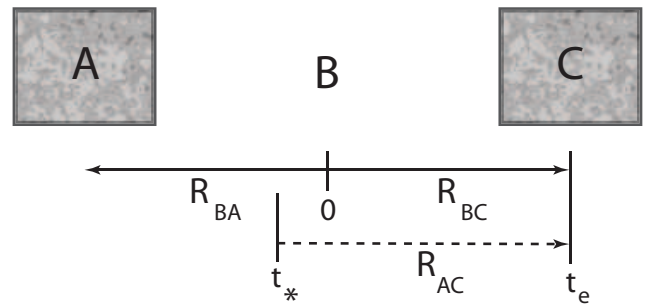}
      \caption{An observer B is receiving light signals from two
patches (A and C) of the CMB. We assume that the light
was emitted by these sources at cosmic time $t_e$, when
their proper distances from the observer were $R_{BA}(t_e)$
and $R_{BC}(t_e)$, respectively. In addition, we assume
that A emitted an earlier light signal at time $t_*$ that reached
C, a proper distance $R_{AC}(t_e)$ away, at time $t_e$.}
\end{figure}

The geodesic Equation~(10) can sometimes be solved analytically,
for example, in $\Lambda$CDM early in the universe's history
when radiation dominated
$\rho$ and $p$ (for which $w=+1/3$). To see how this works in
practice, let us consider a specific scenario (not unlike the simple
thought experiment of \S2), in which observer B is exchanging signals
with patches A and C in the CMB, as shown in Figure~2.
The photons emitted by A and C at time $t_e$ reach the
observer B at a later time $t_0$ (and by symmetry, photons
emitted by B at $t_e$ also reach A and C at $t_0$). Earlier, A emitted a signal at
time $t_*$ ($<t_e$) in order to communicate with C at or before
$t_e$.  Thus, A and C will have been causally connected by
the time their light was emitted at $t_e$ towards B.

Now, by definition, the proper distance between A and C at
time $t_*$ is
\begin{equation}
R_{AC}(t_*)=a(t_*)\int_{t_*}^{t_e}c{dt'\over a(t')}\;.
\end{equation}
We'll think of $t_e$ as the time of recombination, when
matter and radiation separated. Therefore the
universe was radiation dominated up to $t_e$, and
\begin{equation}
a(t)=(2H_0t)^{1/2}\;,
\end{equation}
where $H_0$ is the current value of the Hubble constant 
(Melia \& Abdelqader \cite{MeliaAbdelqader09}).  With this expansion
factor, Equation~(13) is easy to evaluate, yielding the result
\begin{equation}
R_{AC}(t_*)=2ct_*\left[\left({t_e\over t_*}\right)^{1/2}-1\right]\;.
\end{equation}
But in order for the signal emitted by A at $t_*$ to reach C at
$t_e$, we must also have $R_{\gamma e}(t_*)=R_{AC}(t_*)$,
and so for any arbitrary time $t$ ($<t_e$),
\begin{equation}
R_{\gamma e}(t)=2ct\left[\left({t_e\over t}\right)^{1/2}-1\right]\;.
\end{equation}
We have added an ``e"
to the subscript to emphasize that this is the particular geodesic reaching
the observer (in this instance C) at time $t_e$. It is straightforward to
show that Equation~(16) is the solution to Equation~(10),
since in this case $R_{\rm h}(t)=2ct$. The properties of this
geodesic trajectory are representative of all such null paths in
FRW spacetime (Bikwa et al. \cite{Bikwa12}).
Specifically, $R_{\gamma e}(t)\rightarrow 0$
as $t\rightarrow 0$, $R_{\gamma e}(t)\rightarrow 0$ as
(in this particular case) $t\rightarrow t_e$, and $R_{\gamma e}(t)$
has a maximum at $t=t_e/4$, where
$R_{\gamma e}(t_e/4)=ct_e/2$. All of these features are
evident in the photon geodesics shown schematically in Figures~3,
4, and 5.

\section{The Horizon Problem in $\Lambda$CDM}
Figure~3 explains why the observed properties of the CMB
are in conflict with $\Lambda$CDM---a situation known as the ``horizon problem"
in cosmology. In the following discussion, we will find it easier to consider
geodesics drawn from the perspective of someone in patch C. This should
not matter as far as calculated proper distances are concerned, because
the cosmic time $t$ is the same everywhere and if we determine that C
has received a light signal from B during a certain time interval, then it
stands to reason that B will likewise have received a signal from C during
that time.

   \begin{figure}
   \centering
      \includegraphics[angle=0,width=8.3cm]{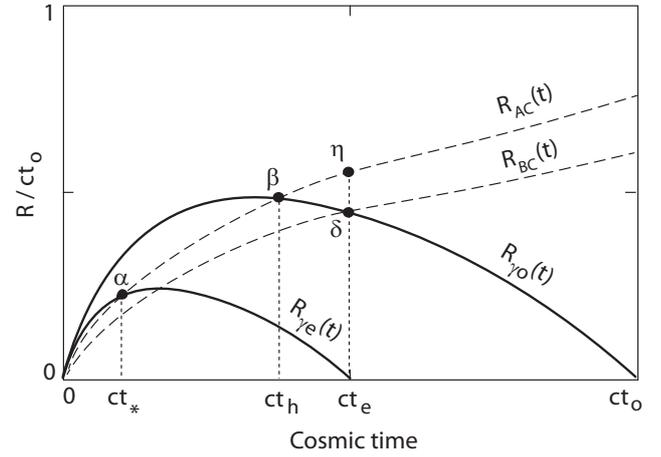}
      \caption{Photon trajectories (solid curves) drawn from the perspective
of an observer in patch C (see Figure~2). The curve $R_{\gamma e}(t)$ is the
null geodesic for light signals arriving at C at time $t_e$ from any
source at proper distance $R_{\rm src}(t)= R_{\gamma e}(t)$
(for $t<t_e$), whereas $R_{\gamma 0}(t)$ is the corresponding
null geodesic for photons arriving at C at time $t_0$ ($>t_e$),
though from sources at $R_{\rm src}(t)=R_{\gamma 0}(t)$
(at any $t<t_0$). The other curves and symbols are defined in the text.}
\end{figure}

Drawn from the perspective of patch C,
Figure~3 shows the photon geodesics (solid curves) reaching the
observer at two specific times. The first of these is the present ($t_0$
in his frame), and some of the light propagating along this path was emitted
by B at time $t_e$, a proper distance $R_{BC}(t_e)=R_{\gamma 0}(t_e)$ away.
The emission point is labeled $\delta$ in the figure. At time $t_0$,
C is also receiving light from patch A, radiated at time $t_h$ ($<t_e$),
from a proper distance $R_{AC}(t_h)=R_{\gamma 0}(t_h)$ away, at
the point labeled $\beta$. We're calling this time $t_h$ because no light
emitted by A at $t>t_h$ could have reached C by $t_0$. In other words,
patch A lies beyond C's current photon horizon for $t>t_h$. (However,
light emitted by A after $t_h$ will become visible to C after $t_0$,
and similarly for light emitted by B after $t_e$.)

For a given cosmology (in this case $\Lambda$CDM), the photon geodesic
labeled $R_{\gamma 0}(t)$ is unique, and if we know the time $t_e$ at
which the light was emitted (say, because we know the redshift at which
the CMB was produced), there is only one point $\delta$ that satisfies the
necessary conditions for C to receive the signal at $t_0$. Our
observations of the CMB allow us to see light emitted by patches A and C
at $t_e\sim 380,000$ years, apparently from equal proper distances on
opposite sides of the sky, so that $R_{AC}(t_e)=2R_{BC}(t_e)$ in
this diagram. Is it possible within the purview of $\Lambda$CDM for us to
find a time $t_*$ to satisfy this proper-distance requirement, while A
and C were still causally connected at time $t_e$?

The answer is yes, but not for a $t_e$ as early as 380,000 years after the
big bang. Writing
\begin{equation}
R_{AC}(t_e)=a(t_e)\int_{t_*}^{t_e}c{dt'\over a(t')}\;,
\end{equation}
we can trivially show for a radiation-dominated universe prior to $t_e$ that
\begin{equation}
R_{AC}(t_e)=2ct_e\left(1-\left[{t_*\over t_e}\right]^{1/2}\right)\;.
\end{equation}
For simplicity, and because the results don't depend on the details of
the calculation, we will next calculate $R_{BC}(t_e)$ under the assumption
that the universe was matter-dominated between $t_e$ and $t_0$.  (The
addition of dark energy would change the proper distance by percentage
points, but far from what is needed to change the answer.) So
we will put
\begin{equation}
R_{BC}(t_e)=a(t_e)\int_{t_e}^{t_0}c{dt'\over a(t')}\;,
\end{equation}
where $a(t)=(3H_0t/2)^{2/3}$ in Einstein-de-Sitter space. Thus
\begin{equation}
R_{BC}(t_e)=3ct_e\left(\left[{t_0\over t_e}\right]^{1/3}-1\right)\;.
\end{equation}
Putting $R_{AC}(t_e)=2R_{BC}(t_e)$ therefore gives
\begin{equation}
\left({t_0\over t_e}\right)^{1/3}={5\over 3}-{2\over 3}\left({t_*\over t_e}\right)^{1/2}\;,
\end{equation}
which shows that $t_e$ is minimized when $t_*\rightarrow 0$ and has the value
$t_e\approx 27t_0/125$, or roughly $1/5$ of $t_0$. Thus, the early
deceleration of the universe following the big bang would not permit patches A and
C to recede from each other quickly enough for us to now see the light they emitted
from opposite sides of the sky as early as $t_e=380,000$ years. We could, however,
see their light if it were emitted within the past 10 billion years or so, {\it even though
they are on opposite sides of the sky}.

This point needs to be emphasized because the situation is similar to that of our
thought experiment in \S2. Just because two patches are moving in opposite
directions from us is not a sufficient reason for them to be out of causal contact.
Nonetheless, the fact that we see the CMB so early in the universe's history
means that either (1) there was some additional acceleration---possibly inflation---that
increased $R_{AC}(t_e)$ to twice $R_{BC}(t_e)$ by $t_e$, or (2) the early
deceleration was not as strong as that in $\Lambda$CDM, or was even
completely absent.  As we shall see in the next section, there is no
such horizon problem in the $R_{\rm h}=ct$ universe.

   \begin{figure}[hp]
   \centering
      \includegraphics[angle=0,width=8.3cm]{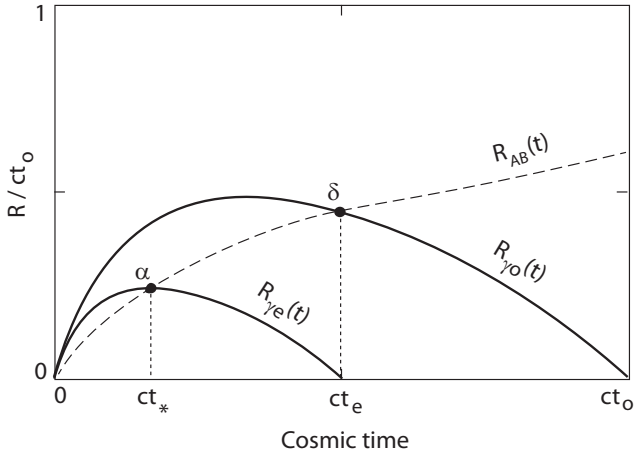}
      \caption{Photon trajectories (solid curves) drawn from the perspective
of an observer in patch B (see Figure~2). By symmetry, the curve $R_{\gamma e}(t)$
is the null geodesic for light signals arriving at A or C at time $t_e$ from B
emitting from a proper distance $R_{AB}(t)= R_{\gamma e}(t)$
(for $t<t_e$), whereas $R_{\gamma 0}(t)$ is the corresponding
null geodesic for photons arriving at B at time $t_0$ ($>t_e$).
The other curves and symbols have the same meanings as in
Figure~3.}
\end{figure}

Before jumping to that case, however, let us consider one more variation
on the scenario we have just explored, one in which the physical conditions
in A and C are equilibrated by virtue of a ``beacon" sending out signals
from B that reach these patches before they emit at $t_e$. This situation
is depicted in Figure~4. Here, we need
\begin{equation}
R_{AB}(t_*)=R_{BC}(t_*)=2ct_*\left(\left[{t_e\over t_*}\right]^{1/2}-1\right)\;,
\end{equation}
but also
\begin{equation}
R_{AB}(t_e)=R_{BC}(t_e)=3ct_e\left(\left[{t_0\over t_e}\right]^{1/3}-1\right)
\end{equation}
(under the same conditions as before). With
\begin{equation}
R_{AB}(t_e)=R_{BC}(t_e)={a(t_e)\over a(t_*)}R_{AB}(t_*)\;,
\end{equation}
it is straightforward to see that the result for $t_e$ is the same as in Figure~3, i.e., the
earliest $t_e$ that would have allowed A and C to be causally connected---and yet appear
on opposite sides of the sky---is roughly $1/5$ of $t_0$ within the context of
$\Lambda$CDM. This is hardly surprising, of course, given the high degree of
symmetry implicit in the FRW metric.

   \begin{figure}
   \centering
      \includegraphics[angle=0,width=8.3cm]{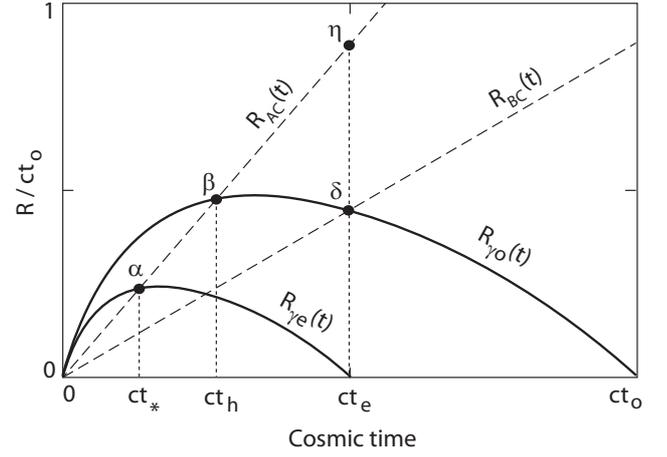}
      \caption{Photon trajectories (solid curves) drawn from the perspective
of an observer in patch C (see Figure~2) for a universe in which
$R_{\rm h}=ct$. All the symbols have the same meanings as those
of Figure~3.}
\end{figure}

\section{No Horizon Problem in the $R_h=ct$ Universe}
The null geodesics and proper distances for the $R_{\rm h}=ct$ universe
are shown in Figure~5, again from the perspective of an observer in patch C
(see Figure~2). The principal difference between this cosmology and $\Lambda$CDM
is that here $a(t)\propto t$ for all cosmic time $t$. Therefore, without necessarily having
to go through the same detailed derivations as before, we can immediately write down
\begin{equation}
R_{AC}(t_e)=ct_e\,\ln(t_e/t_*)\;,
\end{equation}
and
\begin{equation}
R_{BC}(t_e)=ct_e\,\ln(t_0/t_e)\;.
\end{equation}
The question now is ``under what conditions could A and C have been causally connected
at $t_e$, yet with $R_{AC}(t_e)=2R_{BC}(t_e)$ (which means that here on
Earth we would be seeing light emitted by A and C at time $t_e$ from opposite
sides of the sky)?
Clearly, this condition is met when
\begin{equation}
t_*=t_e\,\left({t_e\over t_0}\right)^2\;,
\end{equation}
and just as was true for the thought experiment in \S2, this time $t_*$ may
be small---but it is not zero.

\section{Conclusions}
By carefully tracing null geodesics through the FRW spacetime, we have
affirmed the well-known ``horizon" problem in $\Lambda$CDM, highlighting
its inability to account for the observed properties of the CMB without
some additional early acceleration that would have permitted light
emitted at $t_e\sim 380,000$ years after the big bang to now reach
us from opposite sides of the sky. The inflationary model was invented
largely to offset this flaw in the basic theory.

However, we have also demonstrated that the ``horizon" problem is
not generic to all FRW cosmologies. In particular, we have demonstrated
that the recently introduced $R_{\rm h}=ct$ universe can easily
accommodate a situation in which patches of the CMB, seemingly well
beyond each other's photon horizon on opposite sides of the sky, were nonetheless
in causal contact with each other after the big bang and emitted the radiation
we see today under identical physical conditions. To be clear, these patches
were not really beyond each other's photon horizon, though this mistaken
inference is often made because of the confusion over the meaning
of proper distance $R=a(t)r$. The reason is that $R(t)$ does not represent
the distance $ct$ light has traveled during a time $t$.  One can easily
convince themselves of this by a quick inspection of Equation~(10). Therefore
a simple comparison of the proper distance between two patches of
CMB on opposite sides of the sky with the distance $ct_0$ light could
have traveled since the big bang is not sufficient to demonstrate
whether or not these patches are causally connected.

For various reasons, some of which have been discussed elsewhere
(Melia 2007; Melia \& Shevchuk 2012; Melia 2013; Melia \& Maier 2013;
Wei et al. 2013), we believe that the $R_{\rm h}=ct$ universe
provides a theoretical context for understanding 
the features that have emerged via fits to the data using the 
standard model. In this paper, we have extended the 
argument in favor of the $R_{\rm h}=ct$ universe by 
demonstrating that it can also explain how apparently 
disconnected regions of the CMB on opposite sides of the 
sky were nonetheless in physical equilibrium when they 
produced the microwave radiation we see today. 

Therefore, inflation was not needed to fix a ``horizon"
problem that does not exist in this cosmology.  In the end, this
may turn out to be the most important feature of the $R_{\rm h}=ct$
universe because the inflationary scenario is still not completely
understood, and may not be able to overcome all of
its inherent problems (e.g., the so-called monopole problem;
Guth \& Tye \cite{Guth80}; Einhorn \cite{Einhorn80}).

\begin{acknowledgements}
This research was partially supported by ONR grant N00014-09-C-0032
at the University of Arizona, and by a Miegunyah Fellowship at the
University of Melbourne. I am particularly grateful to Amherst College
for its support through a John Woodruff Simpson Lectureship.
\end{acknowledgements}


\begin{thebibliography}{}

\bibitem[2012] {Bikwa12}
Bikwa, O., Melia, F. \& Shevchuk, A. 2012, MNRAS, 421, 3356

\bibitem[1980] {Einhorn80}
Einhorn, M. B. 1980, Phys. Rev. D, 21, 3295

\bibitem[1981] {Guth81}
Guth, A. H. 1981, Phys. Rev. D, 23, 347

\bibitem[1980] {Guth80}
Guth, A. H. \& Tye, S. 1980, PRL, 44, 631

\bibitem[1982] {Linde82}
Linde, A. 1982, Phys. Lett. B, 108, 389

\bibitem[2007] {Melia07}
Melia, F. 2007, MNRAS, 382, 1917

\bibitem[2012] {Melia12a}
Melia, F. 2012a, Australian Physics, 49, 83 (arXiv:1205.2713)

\bibitem[2012] {Melia12b}
Melia, F. 2012b, AJ, 144, 110

\bibitem[2013] {Melia13}
Melia, F. 2013, ApJ, 764, 72

\bibitem[2013] {MeliaMaier13}
Melia, F. \& Maier, R. S. 2013, MNRAS, submitted

\bibitem[2009] {MeliaAbdelqader09}
Melia, F. \& Abdelqader, M. 2009, IJMP-D, 18, 1889

\bibitem[2012] {MeliaShevchuk12}
Melia, F. \& Shevchuk, A. 2012, MNRAS, 419, 2579

\bibitem[1999] {Perlmutter99}
Perlmutter, S. et al. 1999, ApJ, 517, 565

\bibitem[1998] {Riess98}
Riess, A. G. et al. 1998, AJ, 116, 1009

\bibitem[2003] {Spergel03}
Spergel, D. N. et al. 2003, ApJS, 148, 175

\bibitem[2013] {Wei13}
Wei, Jun-Jie, Wu, X. \& Melia, F. 2013, ApJ, submitted (arXiv:1301.0894)

\end{thebibliography}
\end{document}